# Classification of Flames in Computer Mediated Communications

Nitin
Department of Computer Science, The Peter Kiewit Institute, College of Information Science and Technology, University of Nebraska at Omaha, Omaha-68182-0116, Nebraska, USA

fnunitin@mail.unomaha.edu

Ankush Bansal, Siddhartha Mahadev Sharma, Kapil Kumar, Anuj Aggarwal, Sheenu Goyal, Kanika Choudhary, Kunal Chawla, Kunal Jain and Manav Bhasin
Jaypee University of Information Technology, Waknaghat-173234, via Kandaghat, Dumehar Bani, Himachal Pradesh, India

flaminggroup@googlegroups.com

## ABSTRACT
Computer Mediated Communication (CMC) has brought about a revolution in the way the world communicates with each other. With the increasing number of people, interacting through the internet and the rise of new platforms and technologies has brought together the people from different social, cultural and geographical backgrounds to present their thoughts, ideas and opinions on topics of their interest. CMC has, in some cases, gave users more freedom to express themselves as compared to Face-to-face communication. This has also led to rise in the use of hostile and aggressive language and terminologies uninhibitedly. Since such use of language is detrimental to the discussion process and affects the audience and individuals negatively, efforts are being taken to control them. The research sees the need to understand the concept of flaming and hence attempts to classify them in order to give a better understanding of it. The classification is done on the basis of type of flame content being presented and the Style in which they are presented.

## General Terms
Classification, Computer Mediated Communication, Flames.

## Keywords
Direct flames, indirect flames, straightforward flames, satirical flames, flaming.

## 1. INTRODUCTION
With the advent of computer mediated communications (CMC), human beings have started to communicate with people who are not only their friends, relatives and acquaintances but also with strangers from different social and cultural backgrounds. Information, ideas, thoughts and opinions are exchanged through this medium. The growth in the use of computers to form global networks, which is commonly labeled as the internet, has led to the research being conducted to analyze the benefits and effects of using it as an emerging communication tool. Both modes of internet communication, asynchronous and synchronous, have been able to influence the lives of people around the world. This information is characteristically transmitted to audiences around the world, which shows traits of both written and oral language. The users identified by their 'nickname' adopted by them in a CMC and all messages that transmit appear along with this 'nickname'. At times, the email address of the user is used to identify the user if they have not provided with a desired nickname.

The most common and widely used asynchronous CMC is the e-mail. It is an electronic adaptation of the real life mailing system where a sender leaves a message or 'mail' in a receiver's electronic mailbox (commonly called 'Inbox') and the receiver must open this mailbox and the letter before he can read the messages received [1]. Other more complex and sophisticated example of an asynchronous CMC is 'Usenet Newsgroup' which is an electronic pin-up notice board on which users can post messages regarding certain topics or areas of interest. Other users can read and reply to these messages by opening the notice board and post their own messages in turn. Similar to the e-mail system, there is as such, no real timeline between the users interacting. On the other hand, synchronous CMC provides a real timeline of interactions between users' computers. This is most important feature of synchronous CMC in comparison to asynchronous CMC. The most widespread system, which is an example of synchronous CMC, is Internet Relay chat, or IRC. IRC is a form of CMC, which allows users to 'Chat' by exchanging written messages. They interact in two different ways: By sending their intended messages to a specified user or to all members of the group who are online.

CMC is being largely used by small and large organizations for varied activities such as "group problem solving and forecasting, consensus development, coordination and operation of group projects, sharing ideas and information, and mobilizing organizational action within special forums, interest groups or workforces" [2]. It has been observed that these discussions and conversations sometimes lead to altercations, which are not conducive for the well being of users or organizations per se. Such altercations are also reported in CMC, which has distinct characteristics like that of real life altercations. Such venues of CMC interactions can include examples like forums, blogs, UseNet, IRC chat, micro blogging sites and social networking sites which allow people to come together and discuss on the topics of their interest.

As people started to get more acquainted with CMC, they started to use these heated conversation for their own benefit. Whenever people find that the discussion (mainly debates and arguments) is not going in their favor or they are about to be proved wrong, in that case people use flaming to make the conversation go their way [3]. The use of hostile language in CMC has been reported in most chat forms of synchronous and asynchronous forms of communications. Earlier studies and meta-analysis showed that instances of offensive conduct are highly exaggerated [4] but present scenario of CMC to quite an extent is characterized by intense language, swearing, and hostile communication [3].

## 2. FLAMING

The term "flaming" originates from the early computing community, and *The Hacker's Dictionary* [5] defines it as "to speak rabidly or incessantly on an uninteresting topic or with a patently ridiculous attitude".

Flaming is described as anti-normative hostile and insulting interaction between users [6]. O'Sullivan and Flanagin (2003) define flaming as "a concept emerged from popular discourse surrounding the online community to describe aggressive, hostile, profanity-laced interactions"[7]. Landry (2000) refers to the phenomenon as "uninhibited and aggressive communication" [8]. In general, flaming means to attack with an intention to offend someone through e-mail, posting, commenting or any statement using insults, swearing and hostile, intense language, trolling, etc.[9] .

However, this definition is not considered to be completely appropriate as there are many incidences found where people do not use insulting words or aggressive language directly. Users make use of tools of language such sarcasm, references and figures of speech to flame other user(s) [10, 11]. According to Aiken and Waller, flaming constitutes of "comments intended to offend others. While it is subjective in certain instances, extreme cases of flaming include obscenities and other inappropriate comments." [12]

The term 'Flaming' has also been equated with 'disinhibited' behavior, which is in fact a theorized cause of the action (flaming), rather than the behavior itself [13]. Some researchers have specifically included words such as "electronically" in its definition and used "Computer Mediated" to explain it [14]. Though the term has been adopted from CMC, it has been debatable that the defining flaming only as an online phenomenon is assuming it as technological determinism from a few points of view, confusing the behavior with its theorized causes [7, 13 and 15]. In that respect, several studies have compared flaming on CMC to Face-to-Face communication and have tried to identify the similarities. While some research studies argue that flaming is more apparent in CMC [2], others claimed flaming to be less frequent or rare in both situations [16]. These studies are only considered relevant if flaming by the understood definition is not an online phenomenon [17].The term "flaming" is mainly used in electronic contexts and rarely in non-electronic ones, example the classroom [18].

Flaming has different implication in different scenarios as it has been seen that sometimes the user who resorts to flaming has some advantages whereas in many other case studies the user being flamed has the distinct advantage [19]. Many a times it has been observed that, a user for their redirecting the argument or for forcing one's opinion uses the flaming intentionally. Flaming is used deliberately by the flame sender as a means of diverting the other factions from the original discussion, by sending flames so that they can use it for their own benefit in an attempt to agitate and make the other factions change from present topic of discussion or to remain on a certain topic or point which is preferred by the flame sender.

The essence of this topic is that flaming is a very real phenomenon and to some people, it is even an actual problem. There are reported cases where several distinguished individuals have terminated or abandoned maintaining their weblogs (Online diaries that is open for net users to read and comment on), due to excessive negative or hateful feedback they received on their weblogs [20]. Comments and reader messages to online newspaper and magazine articles have also been strongly criticized for been unnecessarily rude and unparliamentary [21]. Some research suggests the law to provide for the protection of net users against flaming and other misuses of the anonymity inherent on the internet [22, 23].

Some network groups have taken a step in this direction by establishing *netiquette – rules of network usage* – which specially addresses users on how they can write and post their messages. These rules emphasize on obligations for group and self-monitoring to ensure decorum to be maintained throughout discussions by use of correct language, respect for other users and communicative relevance. These netiquette rules differ on different sites. Some offenses are likely to be considered more hostile than others and there is a possibility that what certain groups condemns, others are indifferent to or condones. For example, while posting a certain hostile message directed to another user, it is acceptable on some newsgroups (Examples: alt.fan.warlords and alt.flame), whereas this is usually not allowed and censured by social oriented newsgroups.

## 3. COMPUTER MEDIATED COMMUNICATION VS. FACE-TO-FACE COMMUNICATION

Although, According to researchers, flaming is generally more in computer mediated communication as compare to the face-to-face communication, but such Incidences of flaming are quite low in both [13]. In the same line, initial research comparing CMC and face-to-face communication showed that ''flaming'' and other antisocial interactions were more common in electronic interactions [14].

The reason behind such hostile behavior lies within the fact that in computer mediated communication the identity of the users in most cases remains hidden and salient. This anonymity associated with computer mediation allows the users to be more expressive and less concerned about the results of their actions [24].

Another reason for increased cases of flaming in CMC is the factions that are involved in sending and receiving flames do not share eye contact or physical presence during communication [1]. Hence, much of the expressions, gestures and body language that might normally be present in face-to-face conversations is lost in CMC interactions.

It has been documented that though face-to-face communication and CMC can both be used to discuss any topic; there are marked variations due to the medium of communication. Physical presence plays as an important aspect as an inhibitor to volatile behaviour and as an incentive to be acceptably assertive in the very least. Such "conversations occur in a cooperative environment constantly regulated by mutual adjustment and correction." [25, 26]

## 4. MOTIVATION

The research saw the need to classify flames as it has not been attempted before on the basis of type of content presented and style of language it is presented in. Upon classification, the research aims to help provide with a better understanding of the flaming process and hence would give a better idea for developers to design efficient solutions to the subject matter.

## 5. CLASSIFICATION OF FLAMES

Up until now, much research has been conducted in the field of flaming for the last 25 years but, there have been little attempts to classify flames. The study categorizes on the basis of content Type and Style of flaming. Based on the content of the message or the literature used in the messages, the flames can be classified as "Direct and Intentional Flaming" and "Indirect Flaming". Based on the style or language of conveying the intended meanings, the message can be "Straightforward" or "Satirical".

## 5.1 Direct and Intentional Flaming

Flaming tendencies are noted to be highest when users intentionally use abusive, incendiary and hostile message against another user or faction. This is pre-dominant on different forms of computer mediated communication such. Yet, there are small groups of people who take such steps and use venues like status messages and comments on Social Networking Sites, e-mail message boards, forums, blogs, UseNet, IRC chat etc for flaming. If the mode of communication provides relative anonymity, users tend to flame at higher degrees or are unrestrictive.

Direct flames are characterized as "incendiary messages" [27] and "inflammatory remarks" [28]. Previous research descriptions of flames that can be used to represent direct flaming as "rude or insulting" messages [29], "vicious attacks", "nasty and often profane diatribe" [30], "derisive commentary" [27], "vitriolic online exchanges and poison pen letters" [31].

Direct flaming can also be described as messages, which constitute vicious attacks such as name calling, swearing, insulting on another communicating faction. It is also characterized by the use of rude behavior (may be sexually oriented), offensive, aggressive or hostile language.

This type of flaming can be defined as:

**"Directing hostile, unfriendly, aggressive literature towards other user(s) to show disagreement or oppose their statements or ideologies."**

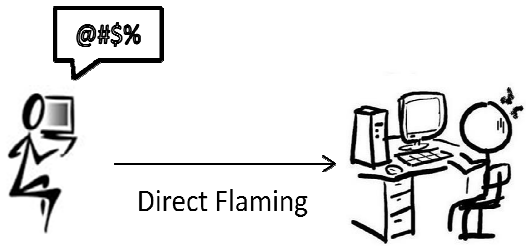

**Fig 1: Direct Flaming**

Such flaming patterns are seldom seen on status messages but are more predominant on discussions on groups or community venues. It is often reported in anonymous synchronous or asynchronous information exchange sites. Even if there is a registration required for posting opinions, pseudo IDs are made to resort to flame any particular person at the same time protecting one's own identity. In most cases, the flame recipient flames back with the same original ID.

Examples of such flaming would be a student flaming a teacher by e-mail, Customers using incendiary language against company officials due to products or services they are not satisfied with. In addition, it is seen that sometimes some offended parties constantly lash out at other users by posting outrageous comments from time to time and by calling them names and use of other methods of insults. Hence, these people end up resorting to personal attacks. Instead of arguing at a professional and mature level regarding the issue at hand and addressing it in a proper way, these flamers resort to choke other participating people through personal attacks. Their sole motive is to heat up the discussions.

One of the most interesting example of Direct flame is the flame war that took place in 1997 on alt.os.linux.slackware newsgroup in which anonymous user ridicules the facts stated in the article in the following manner: "What the f**k do you need so much RAM for? I believe it is not even possible to have that much RAM, or maybe it is, but then you must have a huge simm module of about *8GB RAM. THIS IS CRAZY!!!! YOU DON'T KNOW WHAT THE F**K YOU'RE TALKING ABOUT. Fool!"*

This attracted the response of another anonymous user who replied to the above message with strong use of hostile and abusive language.

An excerpt from the original message: "You swine. You vulgar little maggot. You worthless bag of filth" – and continues clearly highlighting the scorn of the user in the reply [32].

## 5.2 Indirect Flaming

This is generally seen on all forms of computer mediated communication but not as often as direct flaming. Indirect flaming is generally opted for publicizing disagreement or hostility but posted in a language, which can only be understood by the factions involved. Readers of the conversation generally can only note the state of disagreement or recognize the comments as flames but would seldom be able to track references or to whom the flame is intended towards. The users also tend to use sophisticated language in a polite way but can definitely be called as flames when the true meanings of such messages are analyzed for. This characterization can also be used for any hostility intended to third parties in a bilateral conversation, with the intention of initiating aggression against the third party. Another point that can be considered, when people talk about a topic which is specific to their domain for discussion, users without using hostile words, they send flames at each other and it is very difficult to identify whether they are flaming or not. In such cases, only some factions take it as flaming, while others consider it acceptable norms of conversation.

**"Use of hostile, unfriendly and aggressive literature towards users or situations not clearly mentioned, to show disagreement, but with a subtlety that only the factions involved is capable of deriving the true intention of the statement."**

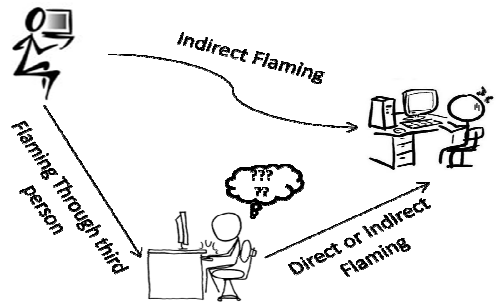

**Fig 2: Indirect Flaming**

Such flaming patterns are seen on status messages of Social Networking Sites, which are made public to all friends. They are also, less frequently used in message boards and e-mails. Due to anonymity offered in most CMC, users prefer Direct

Flaming than this form. Such messages are posted to show strong disagreement in an apparent polite tone.

This term can be used in two different respects: One, the use of indirect references and non-hostile language but still conveying the meaning and intentions. Even though there is little or no abusive language the messages can still be compared to be as incendiary messages having similar effects as that of direct flaming. Secondly, the use of messages or language that initiates a third party to get involved against the flame recipient. The third part could use either direct flames or indirect flames on the recipient.

An example of the secondary usage of indirect flaming, consider a situation where a teacher doesn't give in to the demands of the student, the student contacts the teacher's boss and tells him his side of the story and state the fact that he has been putting an extra effort as compared to his classmates. He also mentions about some of the extra-curricular activities that he had done apart from mainstream academics. Further he flames to the boss that the teacher has been utterly biased against him and that how he had given him a low grade. Finally, he requests him to communicate with the teacher and get his grades revised. The boss then mailed the teacher and probed into the matter. In this case, the flames are intended to the teacher, but are sent via the teacher's boss. This is seen on some moderated message boards and communication sites, or in organizational mailing systems.

These are examples on indirect flames that do not use strong language but still are clear in meaning and intention.

Anonymous user 1: *"I hate you… you're not funny…"*

This was the first message sent by Anonymous user 1 to Anonymous user 2 trying to censure him. As in this message neither hostile nor abusive words are used, so this is indirect flame by anonymous user 1 directed towards anonymous user 2.

Anonymous user 2's Reply: *"I love the fact that you need attention so bad that you had to email me that. That makes me feel happy :) Enjoy your depression."*

The reply to the first user's message is also indirect as the use of language is soft but the intentions are clear. Indirect flames apply the social knowledge of languages and its prevalent usage to convey meanings and intentions [32].

### 5.3 Straightforward Flames

When a user uses straightforward references to people, places or situations without the use of any figure of speech in their messages and with a clear intention of flaming on the topic, it can be termed as straightforward flames or straight flames. This style of flaming is used in tandem with direct or indirect flaming. Since straight flames are more lucid in meaning and the intentions are clear, such messages have higher probability of drawing counter flames by users. Such a style is adopted in CMC where the identity of the flamer is discreet. The users make use of the anonymity of the mode of communication and express more openly, which is a behavior seen frequently on the internet.

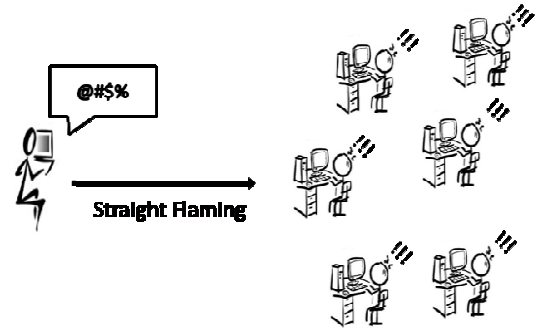

**Fig 3: Straightforward Flaming**

**Table 1. Examples of Different Types of Flaming**

| Types of Flames | Direct | Indirect |
| --- | --- | --- |
| Straight | "Bill, you suck. Euroclean could turn you into an excellent vacuum cleaner. You are an unethical, characterless loser". | "My employers are great but their supervising work force needs replacements big time. They do not seem to work for company profits at all. Such pity". |
| Satirical | An ode to Bill: "you are an awesome guy, Dedicated and passionate, I would love to stand by… Your side, your greatness incarnate. Alas! My words would be true, intention too!! If this were an opposite world where good meant bad and bad good". | "You would be the undisputed king if the world were of donkeys and rats. The world of supervision is of 'wretches and kings'. Alas! Here you're no king". |

### 5.4 Satirical Flames

When a user uses figures of speech or statements which tend to possess alternate derivations aimed at certain factions, places or situations, it can be termed as Satirical Flames. This includes usage of witty language, irony and poetic freedom to convey insults, scorns or even malice. These flames are predominant with users who intended to post incendiary messages but still seem posting normal statements. This style is used along with Direct or Indirect flaming. Since satirical flames are more complicated with references and intentions being vague and enigmatic, responses to such flames are normally enquiring of the details. Generally, the faction about whom the flame was intended chose not respond to such messages when they perceive the true meaning of it. If they do, more often than not, choose to be as enigmatic as the message with their comments.

Let us consider a hypothetical scenario where an employee flames his supervisor of the company in which he is working for, on a public asynchronous micro blogging site.

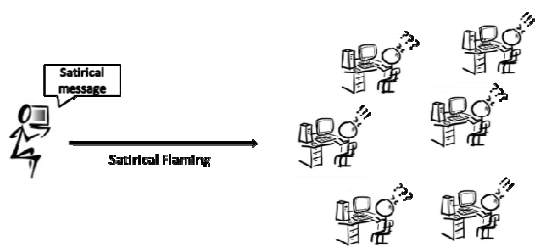

**Fig 4: Satirical Flaming**

## 6. DISCUSSION

The study saw a need to classify flames since it was observed that certain types of flames appeared more frequently than others in certain CMC channels. For example, direct and straightforward flaming is prevalent in anonymous forums or message boards, which provide significant amounts of discretion to the identity of users. The users of such flames at most fear a ban from the site, which some overcome by making alternate accounts to log in. On the other hand, indirect but straightforward or satirical flames are more common on micro blogging sites, social networking sites and other CMC channels where the real identity of the users are partially or completely revealed. It was observed that the discretion of identity of users was directly responsible for type of flames the flamers adopted. The users used more open and offensive tones and language (Direct and straightforward) on sites where their identity was not being compromised while using mild or indirect, enigmatic or satirical flames on sites that linked their online identity to their real life ones. Also, in cases where users of CMC who were less concerned about their behavior, adapted Direct and straightforward modes of flaming to convey a more clear and hostile message.

## 7. CONCLUSION

With the help of different examples and real life flame wars, the study has been able to classify the flames seen in CMC on the basis of type of content presented and the style in which they are presented. Based on type, two types of flames were defined: Direct and Indirect. On the style of language used, the study classified the flames as: Straightforward and Satirical. Direct flames are more prominent and can be easily identified and have a higher negative impact on the audience or the flame receivers than others. Satirical flames are more prominent on sites, which link the user to his/her real identity, such as social networking sites, company mailing lists, etc. All flames can be assigned a type and style based regardless where it is posted.